\documentclass[prl,twocolumn,superscriptaddress,showpacs]{revtex4-2}

\usepackage{amsmath}
\usepackage{units,xspace}
\usepackage{graphicx}
\usepackage{url,hyperref}
\usepackage{multirow}
\usepackage[utf8]{inputenc}

\newcommand{\ifb}{\mbox{fb$^{-1}$}\xspace}

\newcommand{\tev}{\mbox{TeV}\xspace}
\newcommand{\gev}{\mbox{GeV}\xspace}

\newcommand{\gd}{\ensuremath{g_{\textrm D}}\xspace}

\newcommand{\half}{\nicefrac{1}{2}\xspace}
\newcommand{\MAD}{\textsc{MadGraph5}\xspace}
\newcommand{\geant}{\textsc{Geant4}\xspace}
\newcommand{\pf}{\ensuremath{\gamma\gamma}\xspace}

\DeclareGraphicsExtensions{.pdf,.png,.jpg}
\graphicspath{{./}{./fig/}}

\begin{document}


\sloppy


\begin{flushleft}
CERN-EP-2019-039  \\
IFIC/19-09 \\
KCL-PH-TH/2019-17
\end{flushleft}

\newpage

\title{Magnetic Monopole Search with the Full MoEDAL Trapping Detector in $13~\tev$ \\ $pp$ Collisions Interpreted in Photon-Fusion and Drell-Yan Production}

\author{B.~Acharya}
\altaffiliation[Also at ]{International Centre for Theoretical Physics, Trieste, Italy}   
\affiliation{Theoretical Particle Physics \& Cosmology Group, Physics Dept., King's College London, UK}

\author{J.~Alexandre}
\affiliation{Theoretical Particle Physics \& Cosmology Group, Physics Dept., King's College London, UK}

\author{S.~Baines}
\affiliation{Theoretical Particle Physics \& Cosmology Group, Physics Dept., King's College London, UK}

\author{P.~Benes}
\affiliation{IEAP, Czech Technical University in Prague, Czech~Republic}

\author{B.~Bergmann}
\affiliation{IEAP, Czech Technical University in Prague, Czech~Republic}

\author{J.~Bernab\'{e}u}
\affiliation{IFIC, Universitat de Val\`{e}ncia -- CSIC, Valencia, Spain}

\author{A.~Bevan}
\affiliation{School of Physics and Astronomy, Queen Mary University of London, UK}

\author{H.~Branzas}
\affiliation{Institute of Space Science, Bucharest -- M\u{a}gurele, Romania}

\author{M.~Campbell}
\affiliation{Experimental Physics Department, CERN, Geneva, Switzerland}

\author{S.~Cecchini}
\affiliation{INFN, Section of Bologna, Bologna, Italy}

\author{Y.~M.~Cho}
\altaffiliation[Also at ]{Center for Quantum Spacetime, Sogang University, Seoul, Korea} 
\affiliation{Physics Department, Konkuk University, Seoul, Korea}

\author{M.~de~Montigny}
\affiliation{Physics Department, University of Alberta, Edmonton, Alberta, Canada}

\author{A.~De~Roeck}
\affiliation{Experimental Physics Department, CERN, Geneva, Switzerland}

\author{J.~R.~Ellis}
\altaffiliation[Also at ]{National Institute of Chemical Physics \& Biophysics, Tallinn, Estonia} 
\affiliation{Theoretical Particle Physics \& Cosmology Group, Physics Dept., King's College London, UK}
\affiliation{Theoretical Physics Department, CERN, Geneva, Switzerland}

\author{M.~El~Sawy}
\altaffiliation[Also at ]{Department of Physics, Faculty of Science, Beni-Suef University, Beni-Suef, Egypt} 
\altaffiliation{Basic Science Department, Faculty of Engineering, The British University in Egypt, Cairo, Egypt} 
\affiliation{Experimental Physics Department, CERN, Geneva, Switzerland}

\author{M.~Fairbairn}
\affiliation{Theoretical Particle Physics \& Cosmology Group, Physics Dept., King's College London, UK}

\author{D.~Felea}
\affiliation{Institute of Space Science, Bucharest -- M\u{a}gurele, Romania}

\author{M.~Frank}
\affiliation{Department of Physics, Concordia University, Montr\'{e}al, Qu\'{e}bec,  Canada}

\author{J.~Hays}
\affiliation{School of Physics and Astronomy, Queen Mary University of London, UK}

\author{A.~M.~Hirt}
\affiliation{Department of Earth Sciences, Swiss Federal Institute of Technology, Zurich, Switzerland -- Associate member}

\author{J.~Janecek}
\affiliation{IEAP, Czech Technical University in Prague, Czech~Republic}


\author{D.-W.~Kim}
\affiliation{Physics Department, Gangneung-Wonju National University, Gangneung, Republic of Korea}

\author{A.~Korzenev}
\affiliation{D\'epartement de Physique Nucl\'eaire et Corpusculaire, Universit\'e de Gen\`eve, Geneva, Switzerland}

\author{D.~H.~Lacarr\`ere}
\affiliation{Experimental Physics Department, CERN, Geneva, Switzerland}

\author{S.~C.~Lee}
\affiliation{Physics Department, Gangneung-Wonju National University, Gangneung, Republic of Korea}

\author{C.~Leroy}
\affiliation{D\'{e}partement de Physique, Universit\'{e} de Montr\'{e}al, Qu\'{e}bec, Canada}

\author{G.~Levi} 
\affiliation{INFN, Section of Bologna \& Department of Physics \& Astronomy, University of Bologna, Italy}

\author{A.~Lionti}
\affiliation{D\'epartement de Physique Nucl\'eaire et Corpusculaire, Universit\'e de Gen\`eve, Geneva, Switzerland}

\author{J.~Mamuzic}
\affiliation{IFIC, Universitat de Val\`{e}ncia -- CSIC, Valencia, Spain}

\author{A.~Margiotta}
\affiliation{INFN, Section of Bologna \& Department of Physics \& Astronomy, University of Bologna, Italy}

\author{N.~Mauri}
\affiliation{INFN, Section of Bologna, Bologna, Italy}

\author{N.~E.~Mavromatos}
\affiliation{Theoretical Particle Physics \& Cosmology Group, Physics Dept., King's College London, UK}

\author{P.~Mermod}
\affiliation{D\'epartement de Physique Nucl\'eaire et Corpusculaire, Universit\'e de Gen\`eve, Geneva, Switzerland}

\author{M.~Mieskolainen}
\affiliation{Physics Department, University of Helsinki, Helsinki, Finland}

\author{L.~Millward}
\affiliation{School of Physics and Astronomy, Queen Mary University of London, UK}

\author{V.~A.~Mitsou}
\email[Corresponding author: ]{vasiliki.mitsou@ific.uv.es}
\affiliation{IFIC, Universitat de Val\`{e}ncia -- CSIC, Valencia, Spain}

\author{R.~Orava}
\affiliation{Physics Department, University of Helsinki, Helsinki, Finland}

\author{I.~Ostrovskiy}
\affiliation{Department of Physics and Astronomy, University of Alabama, Tuscaloosa, Alabama, USA}

\author{J.~Papavassiliou}
\affiliation{IFIC, Universitat de Val\`{e}ncia -- CSIC, Valencia, Spain}

\author{B.~Parker}
\affiliation{Institute for Research in Schools, Canterbury, UK}

\author{L.~Patrizii}
\affiliation{INFN, Section of Bologna, Bologna, Italy}

\author{G.~E.~P\u{a}v\u{a}la\c{s}}
\affiliation{Institute of Space Science, Bucharest -- M\u{a}gurele, Romania}

\author{J.~L.~Pinfold}
\affiliation{Physics Department, University of Alberta, Edmonton, Alberta, Canada}

\author{V.~Popa}
\affiliation{Institute of Space Science, Bucharest -- M\u{a}gurele, Romania}

\author{M.~Pozzato}
\affiliation{INFN, Section of Bologna, Bologna, Italy}

\author{S.~Pospisil}
\affiliation{IEAP, Czech Technical University in Prague, Czech~Republic}

\author{A.~Rajantie}
\affiliation{Department of Physics, Imperial College London, UK}

\author{R.~Ruiz~de~Austri}
\affiliation{IFIC, Universitat de Val\`{e}ncia -- CSIC, Valencia, Spain}

\author{Z.~Sahnoun}
\altaffiliation[Also at ]{Centre for Astronomy, Astrophysics and Geophysics, Algiers, Algeria}
\affiliation{INFN, Section of Bologna, Bologna, Italy}

\author{M.~Sakellariadou}
\affiliation{Theoretical Particle Physics \& Cosmology Group, Physics Dept., King's College London, UK}

\author{A.~Santra}
\affiliation{IFIC, Universitat de Val\`{e}ncia -- CSIC, Valencia, Spain}

\author{S.~Sarkar}
\affiliation{Theoretical Particle Physics \& Cosmology Group, Physics Dept., King's College London, UK}

\author{G.~Semenoff}
\affiliation{Department of Physics, University of British Columbia, Vancouver, British Columbia, Canada}

\author{A.~Shaa}
\affiliation{Physics Department, University of Alberta, Edmonton, Alberta, Canada}

\author{G.~Sirri}
\affiliation{INFN, Section of Bologna, Bologna, Italy}

\author{K.~Sliwa}
\affiliation{Department of Physics and Astronomy, Tufts University, Medford, Massachusetts, USA}

\author{R.~Soluk}
\affiliation{Physics Department, University of Alberta, Edmonton, Alberta, Canada}

\author{M.~Spurio}
\affiliation{INFN, Section of Bologna \& Department of Physics \& Astronomy, University of Bologna, Italy}

\author{M.~Staelens}
\affiliation{Physics Department, University of Alberta, Edmonton, Alberta, Canada}

\author{M.~Suk}
\affiliation{IEAP, Czech Technical University in Prague, Czech~Republic}

\author{M.~Tenti}
\affiliation{INFN, CNAF, Bologna, Italy}

\author{V.~Togo}
\affiliation{INFN, Section of Bologna, Bologna, Italy}

\author{J.~A.~Tuszy\'{n}ski}
\affiliation{Physics Department, University of Alberta, Edmonton, Alberta, Canada}

\author{V.~Vento}
\affiliation{IFIC, Universitat de Val\`{e}ncia -- CSIC, Valencia, Spain}

\author{O.~Vives}
\affiliation{IFIC, Universitat de Val\`{e}ncia -- CSIC, Valencia, Spain}

\author{Z.~Vykydal}
\affiliation{IEAP, Czech Technical University in Prague, Czech~Republic}

\author{A.~Wall}
\affiliation{Department of Physics and Astronomy, University of Alabama, Tuscaloosa, Alabama, USA}

\author{I.~S.~Zgura}
\affiliation{Institute of Space Science, Bucharest -- M\u{a}gurele, Romania}

\collaboration{THE MoEDAL COLLABORATION}
\noaffiliation

\date{July 9, 2019}

\begin{abstract}
MoEDAL is designed to identify new physics in the form of stable or pseudostable highly ionizing particles produced in high-energy Large Hadron Collider (LHC) collisions. Here we update our previous search for magnetic monopoles in Run~2 using the full trapping detector with almost four times more material and almost twice more integrated luminosity. For the first time at the LHC, the data were interpreted in terms of photon-fusion monopole direct production in addition to the Drell-Yan-like mechanism. The MoEDAL trapping detector, consisting of 794~kg of aluminum samples installed in the forward and lateral regions, was exposed to 4.0~\ifb of 13~\tev proton-proton collisions at the LHCb interaction point and analyzed by searching for induced persistent currents after passage through a superconducting magnetometer. Magnetic charges equal to or above the Dirac charge are excluded in all samples. Monopole spins~0, \half and 1 are considered and both velocity-independent and \mbox{-dependent} couplings are assumed. This search provides the best current laboratory constraints for monopoles with magnetic charges ranging from two to five times the Dirac charge.
\end{abstract}

\pacs{14.80.Hv, 13.85.Rm, 29.20.db, 29.40.Cs}

\maketitle


The existence of a magnetically charged particle would add symmetry to Maxwell's equations and explain why electric charge is quantized in Nature, as shown by Dirac in 1931~\cite{Dirac:1931kp}. Dirac predicted the fundamental magnetic charge number (or Dirac charge) to be $\frac{e}{2\alpha_{\rm em}}\simeq 68.5e$ where $e$ is the proton charge and $\alpha_{\rm em}$ is the fine-structure constant. Consequently, in SI units, the magnetic charge can be written in terms of the dimensionless quantity \gd as $q_{\rm m} = n \gd ec$ where $n$ is an integer number and $c$ is the speed of light in vacuum. Because \gd is large, a fast monopole can induce ionization in matter thousands of times higher than a particle carrying the elementary electric charge. 

It has subsequently been shown by 't Hooft and Polyakov that the existence of the monopole as a topological soliton is a prediction of theories of the unification of forces~\cite{tHooft:1974kcl,Polyakov:1974ek,Scott:1979zu,Preskill:1984gd}. For a unification scale of $10^{16}~\gev$ such monopoles would have a mass $M$ in the range $10^{17}-10^{18}~\gev$. In unification theories involving a number of symmetry-breaking scales~\cite{Lazarides:1980va,Kirkman:1981ck,Kephart:2001ix} monopoles of much lower mass can arise, although still beyond the reach of the Large Hadron Collider (LHC). However, an electroweak monopole has been proposed~\cite{Cho:1996qd,Cho:2013vba,Ellis:2016glu,Cho:2016npz} that is a hybrid of the Dirac and 't Hooft--Polyakov monopoles~\cite{tHooft:1974kcl,Polyakov:1974ek} with a mass potentially accessible at the LHC and a minimum magnetic charge 2\gd, underlining the importance of searching for large magnetic charges at the LHC. 

There have been extensive searches for monopole relics from the early Universe in cosmic rays and in materials~\cite{Burdin:2014xma,Patrizii:2015uea}. The LHC has a comprehensive monopole search program using various techniques devised to probe TeV-scale monopole masses for the first time~\cite{DeRoeck:2011aa,Acharya:2014nyr,Aad:2012qi}. The results obtained by MoEDAL using 8~\tev $pp$ collisions allowed the previous LHC constraints on monopole pair production~\cite{Aad:2015kta} to be improved to provide limits on monopoles with $|g|\le 3 \gd$ and $M\le 3500$~\gev~\cite{MoEDAL:2016jlb}, where $g=q_{\mathrm m}ec$. At 13~\tev LHC energies, MoEDAL extended the limits to $|g|\le 5 \gd$ and masses up to $1790~\gev$ assuming Drell-Yan (DY) production~\cite{Acharya:2016ukt,Acharya:2017cio}.

\begin{figure}[htb]
\begin{center}
  \includegraphics[width=\linewidth]{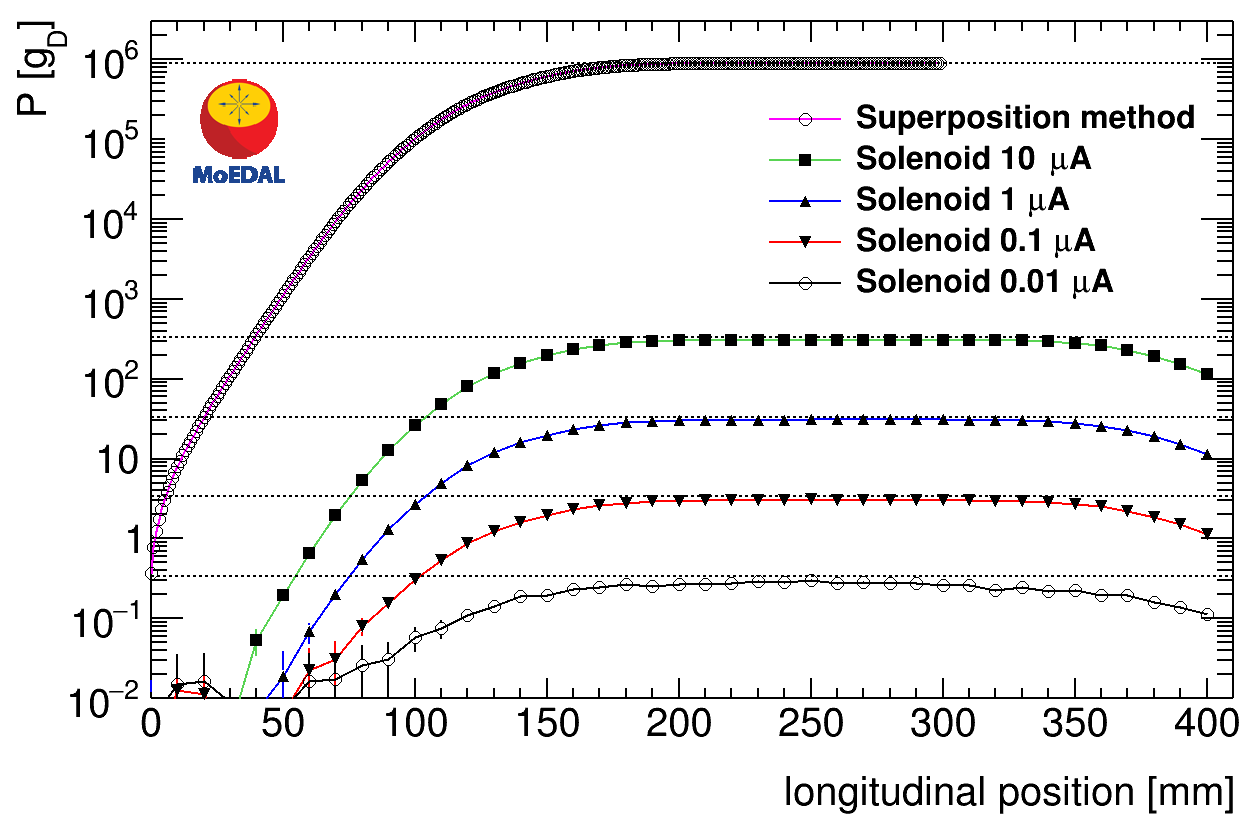}
  \caption{Results of the calibration measurements with the superposition method using a magnetic dipole sample, and the solenoid method with $P=32.4\gd/\mu$A and various currents. The dashed lines represent the expected plateau values in units of Dirac charge. The calibration constant is tuned using the measurement from the superposition method. }
\label{fig:SQUIDcalib}
\end{center}
\end{figure}

In addition to the forward part used in previous analyses~\cite{Acharya:2016ukt,Acharya:2017cio}, the exposed Magnetic Monopole Trapper (MMT) volume analyzed here includes lateral components increasing the total aluminum mass to 794~kg; a schematic view is provided in the Supplemental Material~\cite{supplemental}. All 2400 trapping detector samples were scanned in 2018 with a dc SQUID long-core magnetometer (2G Enterprises Model 755) installed at the Laboratory for Natural Magnetism at ETH Zurich. The measured magnetometer response is translated into a magnetic pole $P$ in units of Dirac charge by multiplying by a calibration constant $C$. Calibration was performed using two independent methods, described in more detail in Ref.~\cite{DeRoeck:2012wua}. The first method adds measurements performed at 1~mm intervals using a dipole sample of known magnetic moment $\mu=2.98\times 10^{-6}$ A\,m$^2$ to obtain the response of a single magnetic pole of strength $P=9.03\times 10^5\gd$, based on the superposition principle. The second method measures directly the effect of a magnetic pole of known strength using a long thin solenoid providing $P=32.4\gd/\mu$A for various currents ranging from 0.01 to 10~$\mu$A. The results of the calibration measurements, with the calibration constant obtained from the first method, are shown in Fig.~\ref{fig:SQUIDcalib}. The two methods agree within 10\%, which can be considered as the calibration uncertainty in the pole strength. The magnetometer response is measured to be linear and charge symmetric in a range corresponding to $0.3-300\gd$. The plateau value of the calibration dipole sample was remeasured regularly during the campaign and was found to be stable to within less than 1\%. 

\begin{figure}[htb]
\begin{center}
  \includegraphics[width=\linewidth]{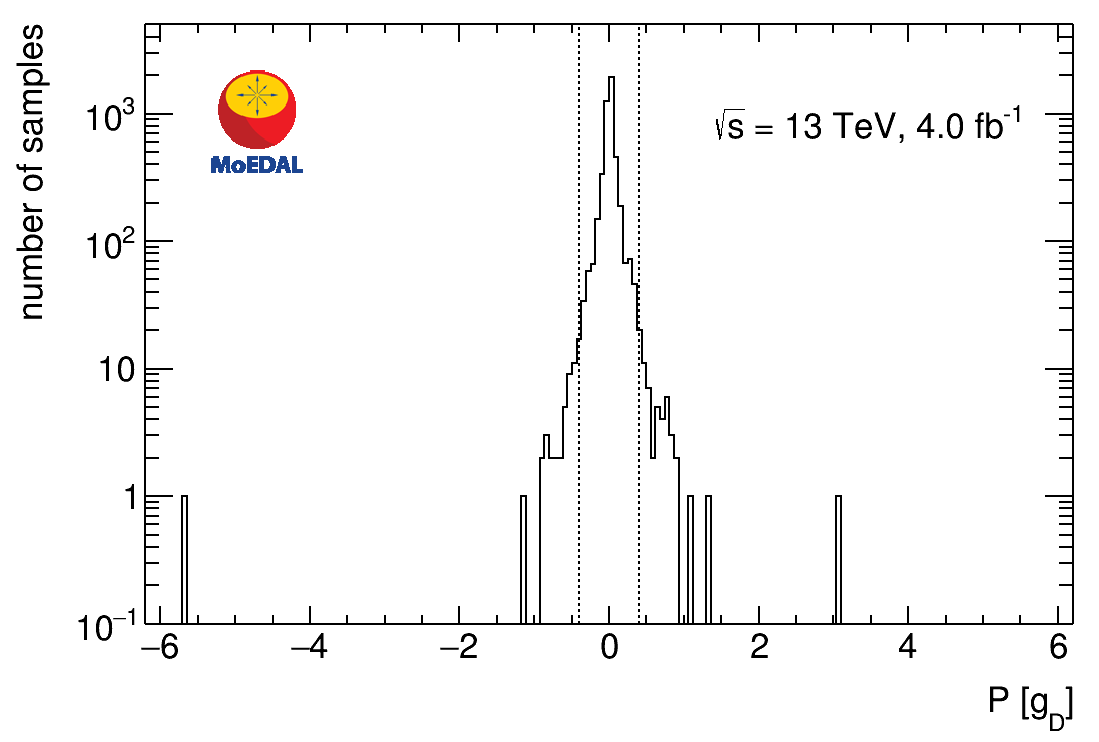}
  \caption{Magnetic pole strength (in units of Dirac charge) measured in the 2400 aluminum samples of the MoEDAL trapping detector exposed to 13~\tev collisions in 2015$-$2017, with every sample scanned twice.}
\label{fig:SQUIDmeasurements}
\end{center}
\end{figure}

\begin{figure}[htb]
\begin{center}
  \includegraphics[width=\linewidth]{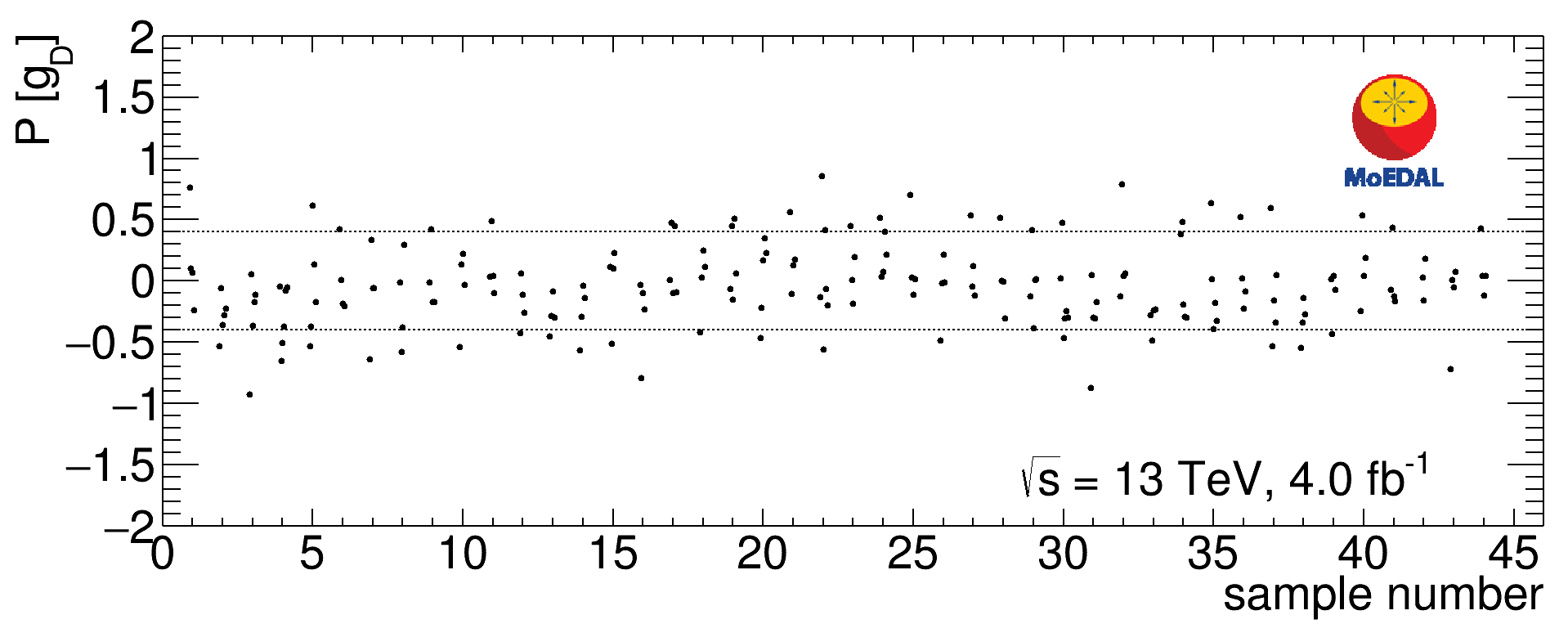}
  \includegraphics[width=\linewidth]{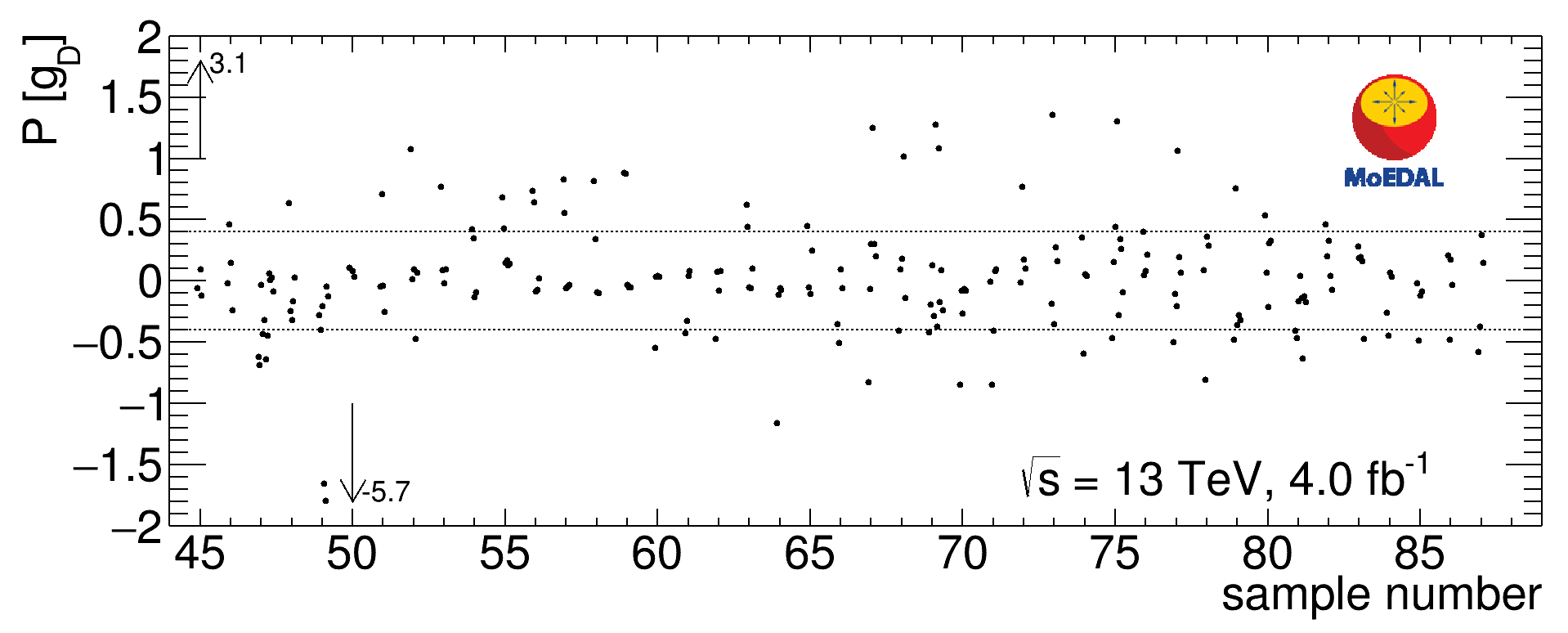}  
  \caption{Results of multiple pole strength measurements (in units of Dirac charge) for the 87~candidate samples for which at least one of the two first measurement values was above the threshold $|g|>0.4\gd$. More values are observed below threshold than above threshold for all of them, excluding the presence of a monopole with $|g|\geq \gd$.}
\label{fig:SQUIDcandidates}
\end{center}
\end{figure}

Samples were placed on a carbon-fiber movable conveyer tray for transport through the sensing region of the magnetometer, three at a time, separated by a distance of 46~cm. The transport speed was set to the minimum available of 2.54~cm/s, as it was found in previous studies that the frequency and magnitude of possible spurious offsets increased with speed~\cite{Acharya:2017cio}. The magnetic charge contained in a sample is measured as a persistent current in the superconducting coil surrounding the transport axis. This is defined as the difference between the currents measured after ($I_2$) and before ($I_1$) passage of a sample through the sensing coil, after adjustment for the corresponding contributions of the empty tray $I^{\text{tray}}_2$ and $I^{\text{tray}}_1$. Expressed in Dirac charges, the magnetic pole strength contained in a sample is thus calculated as $P=C\left[(I_2-I_1)-(I^{\text{tray}}_2-I^{\text{tray}}_1)\right]$, where $C$ is the calibration constant. All samples were scanned twice, with the resulting pole strengths shown in Fig.~\ref{fig:SQUIDmeasurements}. The samples are not subject to any external magnetic field when passed through the superconducting loop that could possibly unbind a monopole from the material. The observed outliers may be due to spurious flux jumps occurring by ferromagnetic impurities in the sample, noise currents in the SQUID feedback loop and other known instrumental and environmental factors~\cite{MoEDAL:2016jlb}. Whenever the measured pole strength differed from zero by more than $0.4\gd$ in either of the two measurements, the sample was considered a candidate. This procedure strongly reduces the possibility of false negatives. A total of 87 candidate samples were thus identified. A sample containing a genuine monopole would consistently yield the same nonzero value for repeated measurements, while values repeatedly consistent with zero would be measured when no monopole is present. The candidates were scanned repeatedly and it was found that the majority of the measured pole strengths for each candidate lay below the threshold of $0.4\gd$, as shown in Fig.~\ref{fig:SQUIDcandidates}. Using the multiple candidate measurements to model the probability distribution of pole strength values, in the worst case in which one misses a monopole three times out of five measurements, an estimated false-negative probability of less than 0.2\% is obtained for magnetic charges of 1\gd. We are thus able to exclude the presence of a monopole with $|g|\geq \gd$ in all samples, including all candidates.

The trapping detector acceptance, defined as the probability that a monopole of given mass, charge, energy and direction would end its trajectory inside the trapping volume, is determined from the knowledge of the material traversed by the monopole~\cite{Alves:2008zz,MoEDAL:2016jlb} and the ionization energy loss of monopoles when they go through matter~\cite{Ahlen:1978jy,Ahlen:1980xr,Ahlen:1982mx,Cecchini:2016vrw}, implemented in a simulation based on \geant~\cite{Allison:2006ve}. For a given mass and charge, the pair-production model determines the kinematics and the overall trapping acceptance obtained. The uncertainty in the acceptance is dominated by uncertainties in the material description~\cite{MoEDAL:2016jlb,Acharya:2016ukt,Acharya:2017cio}. This contribution is estimated by performing simulations with hypothetical material conservatively added and removed from the nominal geometry model. 

\begin{figure}[htb]
\begin{center}
  \includegraphics[width=0.45\linewidth]{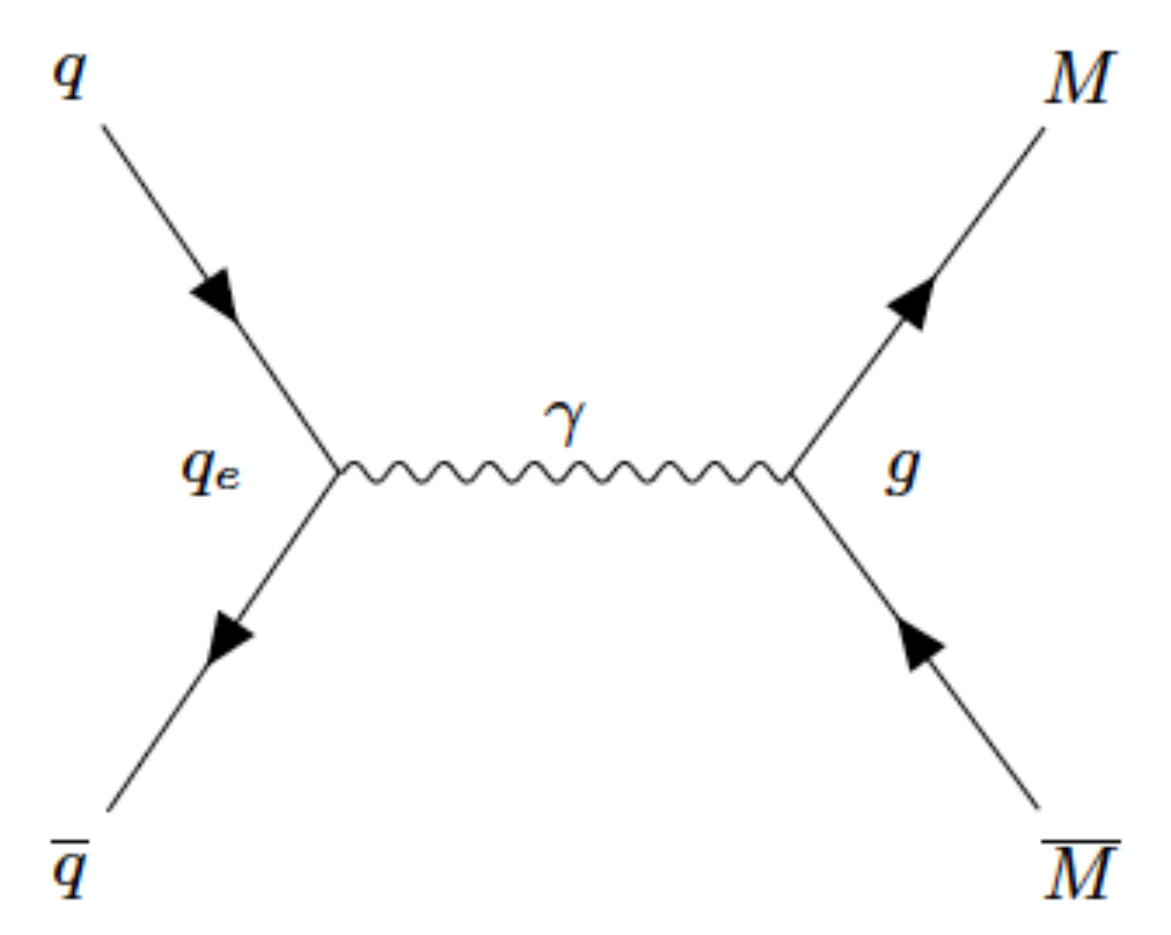}
  \includegraphics[width=0.45\linewidth]{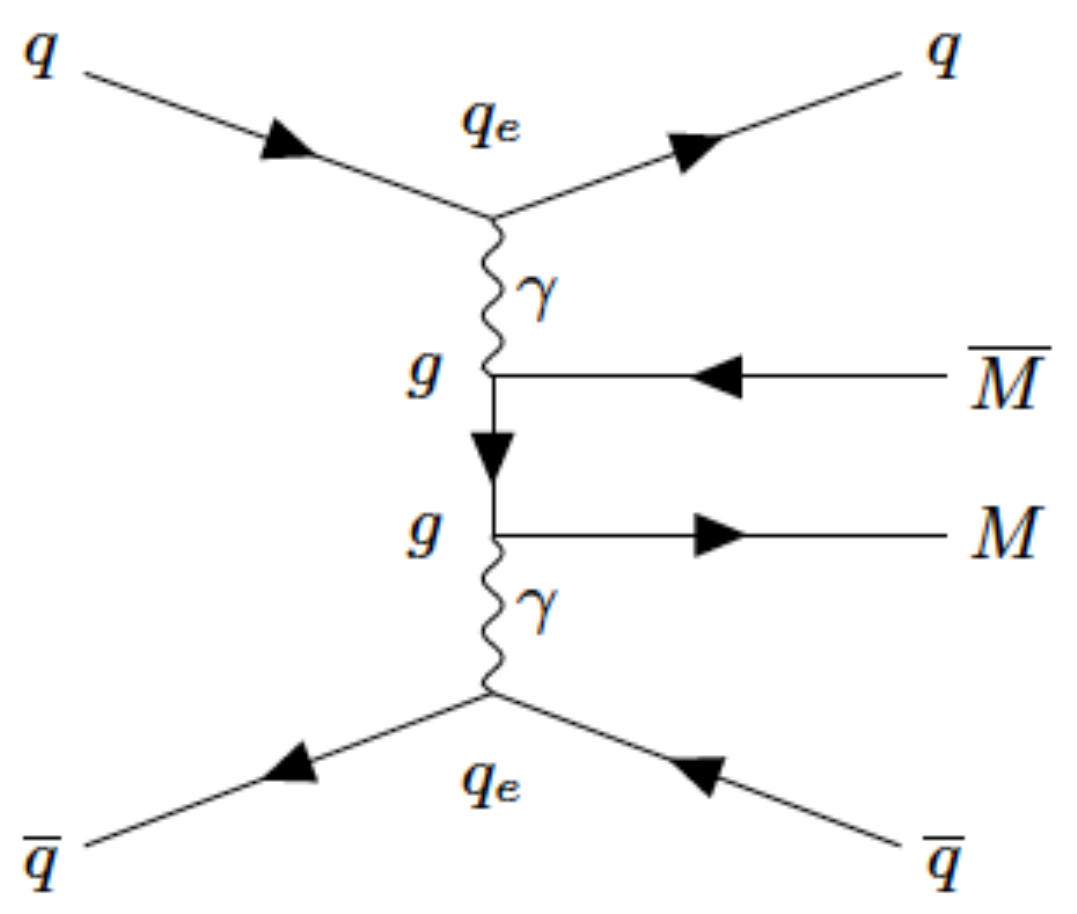}
  \caption{Feynman-like diagrams for monopole pair direct production at leading order via the Drell-Yan (left) and photon-fusion (right) processes at the LHC. For scalar and vector monopoles a four-vertex diagram is also added~\cite{Baines:2018ltl}.}
\label{fig:diagrams}
\end{center}
\end{figure}

A DY mechanism (Fig.~\ref{fig:diagrams}, left) is traditionally employed in searches as it provides a simple model of monopole pair production~\cite{Aad:2012qi,Aad:2015kta,MoEDAL:2016jlb,Acharya:2016ukt,Acharya:2017cio}. In the interpretation of the present search, photon fusion (\pf) (Fig.~\ref{fig:diagrams}, right)~\cite{Baines:2018ltl} is considered in addition to DY for the first time at the LHC, having previously only been used in a collider search for direct  monopole production at the H1 experiment at HERA~\cite{Aktas:2004qd}. An earlier D0 analysis of diphoton events at Tevatron used virtual-monopole production via photon fusion to set limits on the monopole mass~\cite{Abbott:1998mw}.

The different direct production mechanisms, DY and \pf, imply different kinematical distributions, as shown in the Supplemental Material~\cite{supplemental}. However, due to the considerably higher cross section for \pf over most of the spin and mass range~\cite{Baines:2018ltl}, the \pf mechanism is dominant for setting mass bounds. For both processes the cross sections are computed using the Feynman-like diagrams shown in Fig.~\ref{fig:diagrams}, although the large monopole coupling to the photon places such calculations in the nonperturbative regime. A proposal involving the thermal Schwinger production of monopoles in heavy-ion collisions~\cite{Schwinger:1951nm}, which does not rely on perturbation theory,  overcomes these limitations~\cite{Gould:2017fve,Gould:2019myj}. Here the subsequent combination of the production processes implies merely summing the total cross sections computed from these leading-order diagrams, respecting at the same time the different kinematics. No interference terms are considered. 

As in the previous MoEDAL MMT analysis~\cite{Acharya:2017cio}, monopoles of spins 0, \half and~1 are considered, with the values of the monopole magnetic moment assumed to be zero for spin~\half and one for spin~1, i.e., equal to the Standard Model values for particles with these spins. Models were generated in \MAD~\cite{Alwall:2014hca} using the Universal \textsc{Feyn}\textsc{Rules} Output described in Ref.~\cite{Baines:2018ltl}. We used tree-level diagrams and the parton distribution functions \texttt{NNPDF23}~\cite{Ball:2012cx} and \texttt{LUXqed}~\cite{Manohar:2016nzj} for the DY and \pf production processes, respectively. \texttt{LUXqed} is determined in a model-independent manner using $ep$ scattering data and is the most accurate photon PDF available to date. In addition to a pointlike QED coupling, we have also considered a modified photon-monopole coupling in which $g$ is substituted by $\beta g$ with $\beta = \sqrt{1 - \frac{4M^2}{s}}$ (where $M$ is the mass of the monopole and $\sqrt{s}$ is the invariant mass of the monopole-antimonopole pair), as in the previous MoEDAL analysis~\cite{Acharya:2017cio}. This ``$\beta$-dependent coupling'' illustrates the range of theoretical uncertainties in monopole dynamics close to threshold. Moreover, in the case of spin-\half and spin-1 monopoles, together with the introduction of a magnetic-moment phenomenological parameter $\kappa$, the $\beta$-dependent coupling may lead to a perturbative treatment of the cross-section calculation~\cite{Baines:2018ltl}. 

\begin{figure*}[!htb]
  \begin{center}
    \includegraphics[width=0.495\linewidth]{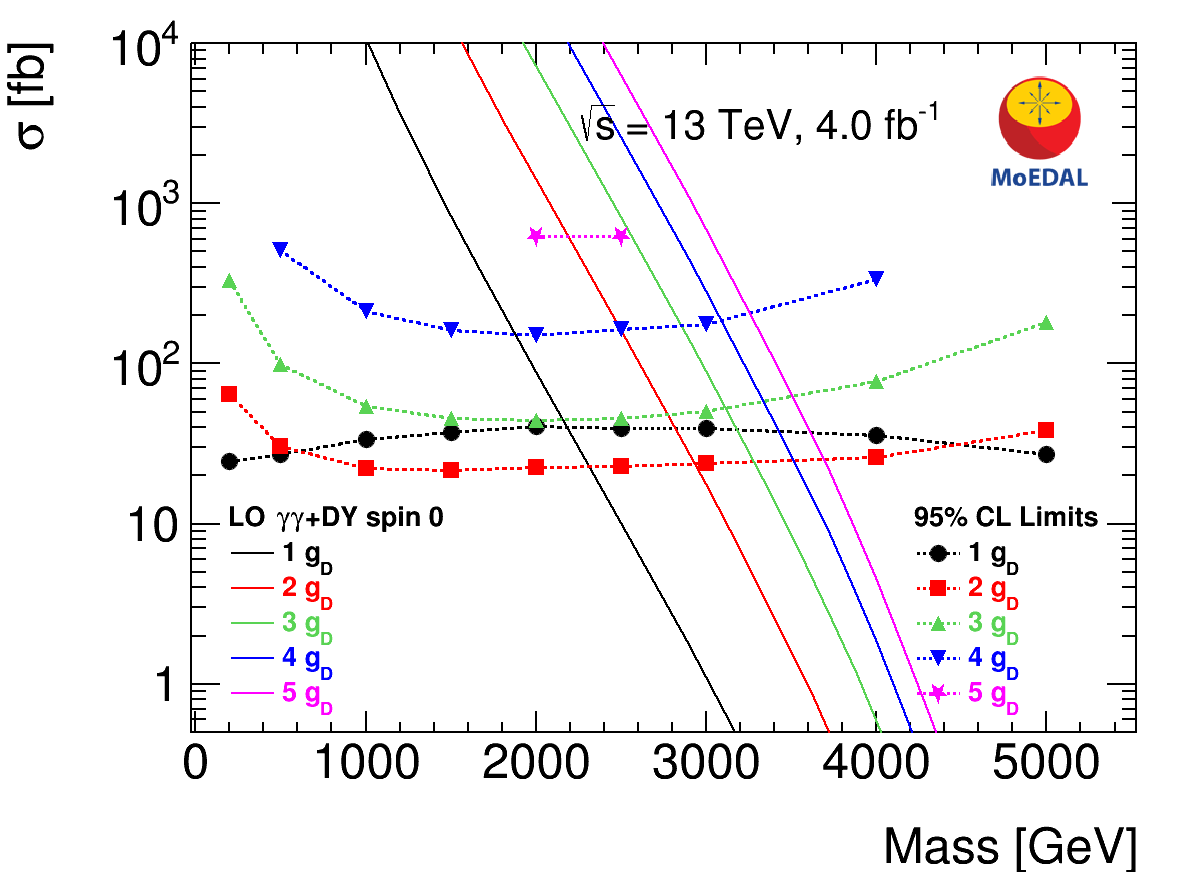}  
    \includegraphics[width=0.495\linewidth]{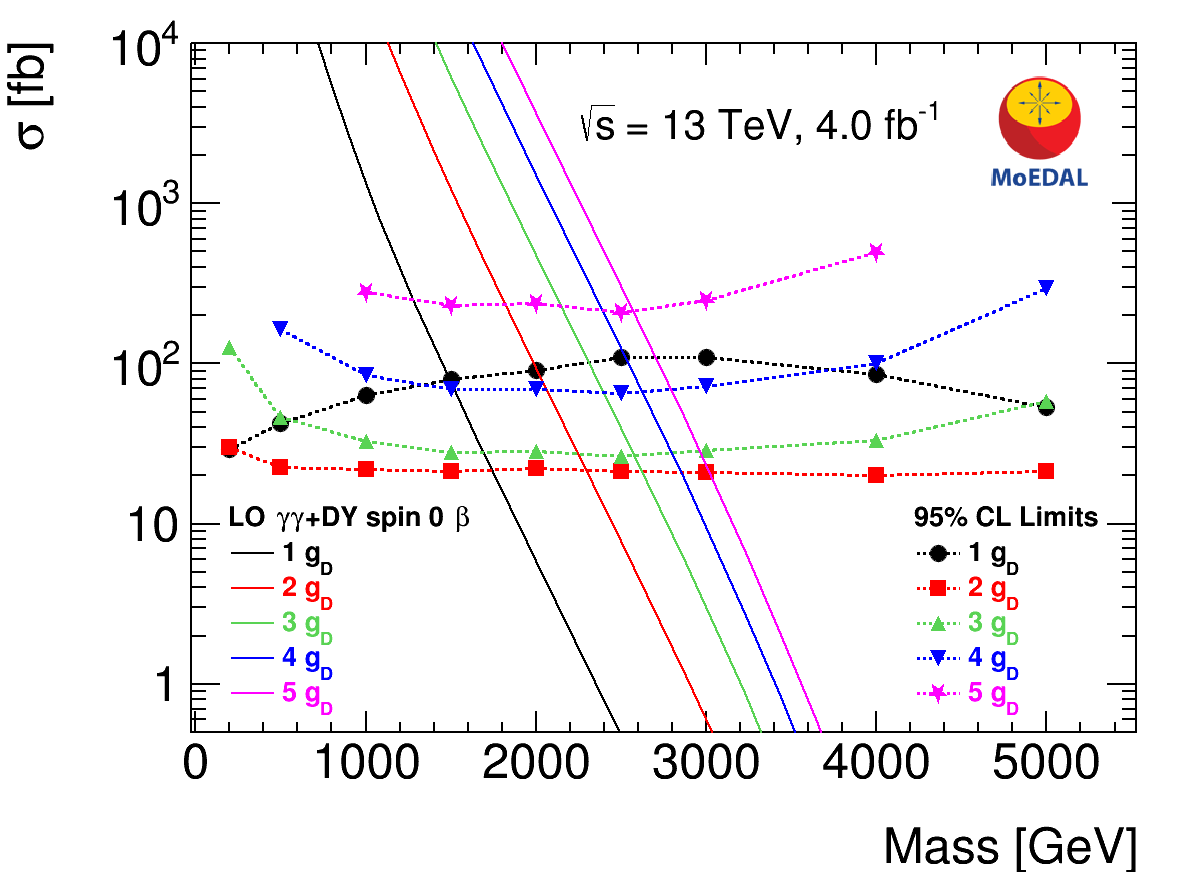}  
    \includegraphics[width=0.495\linewidth]{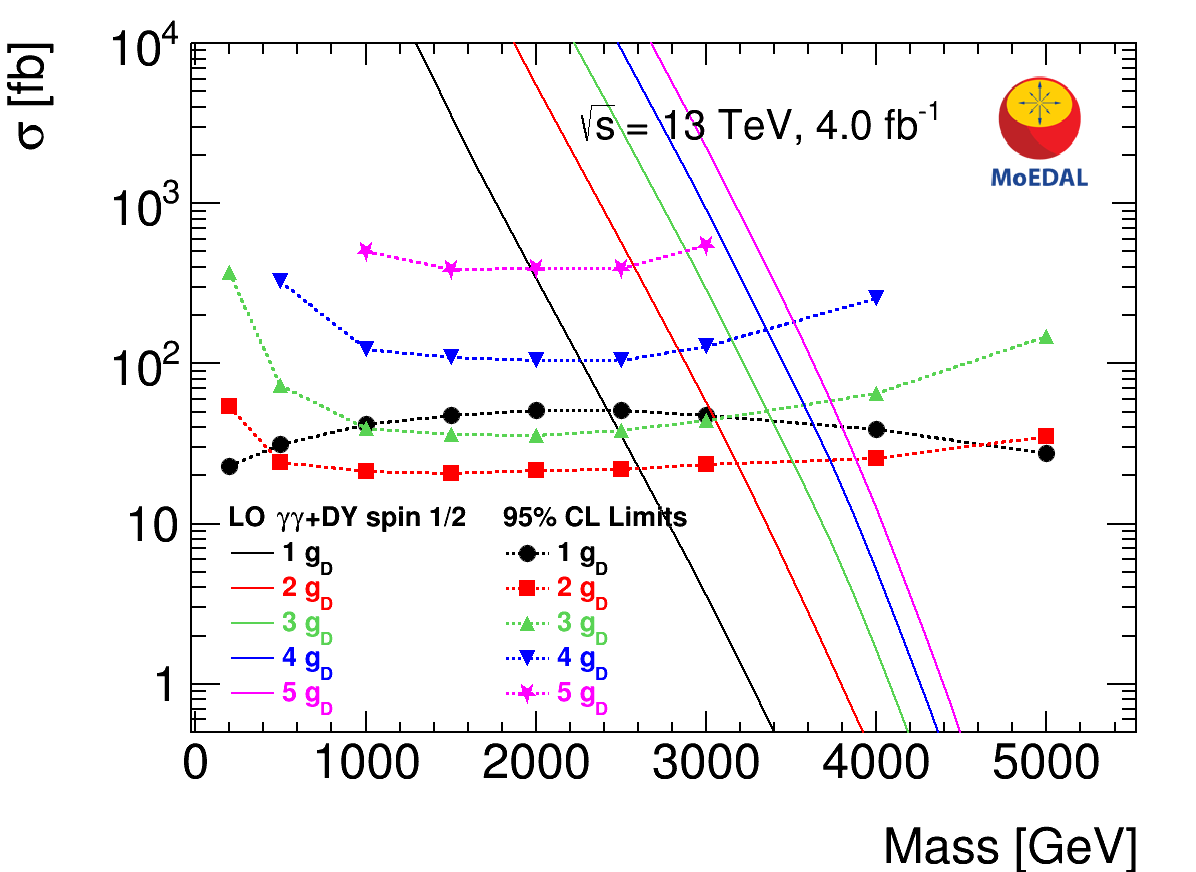}  
    \includegraphics[width=0.495\linewidth]{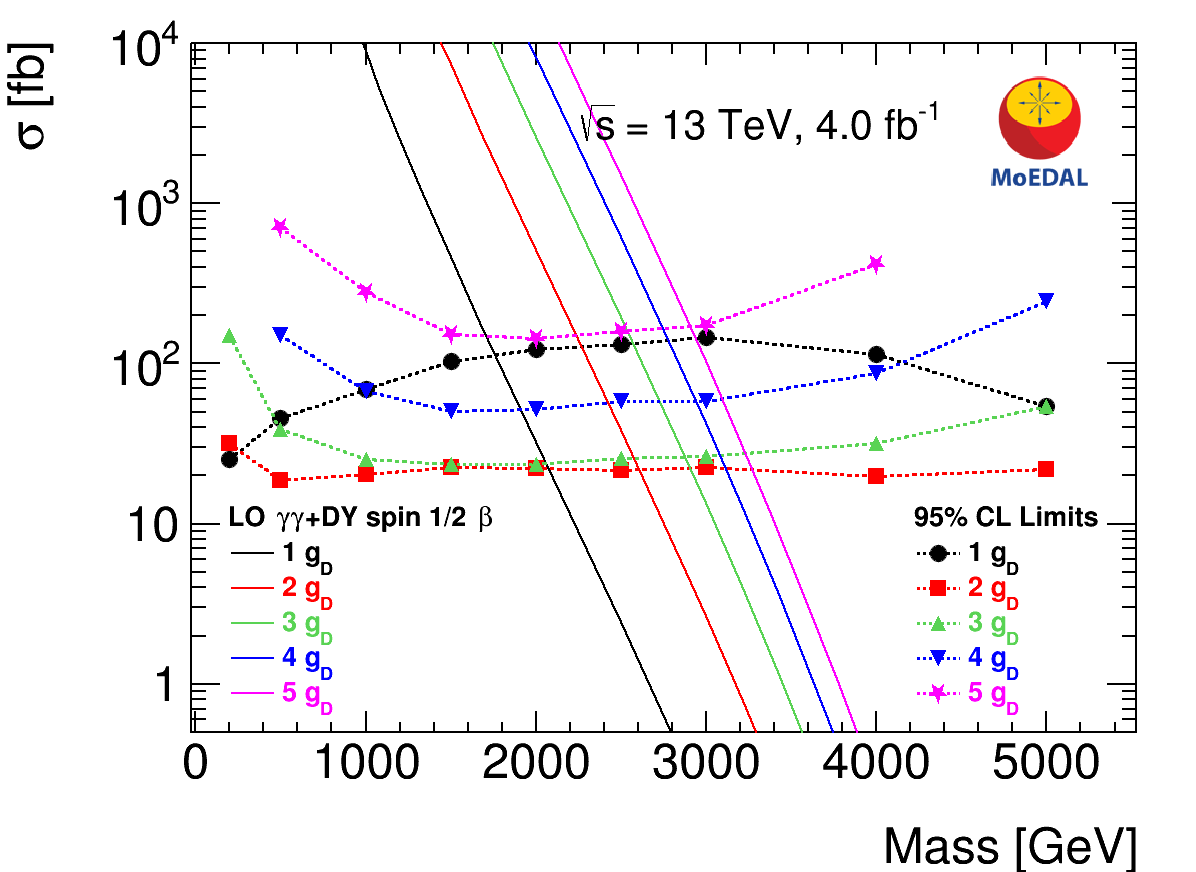}  
    \includegraphics[width=0.495\linewidth]{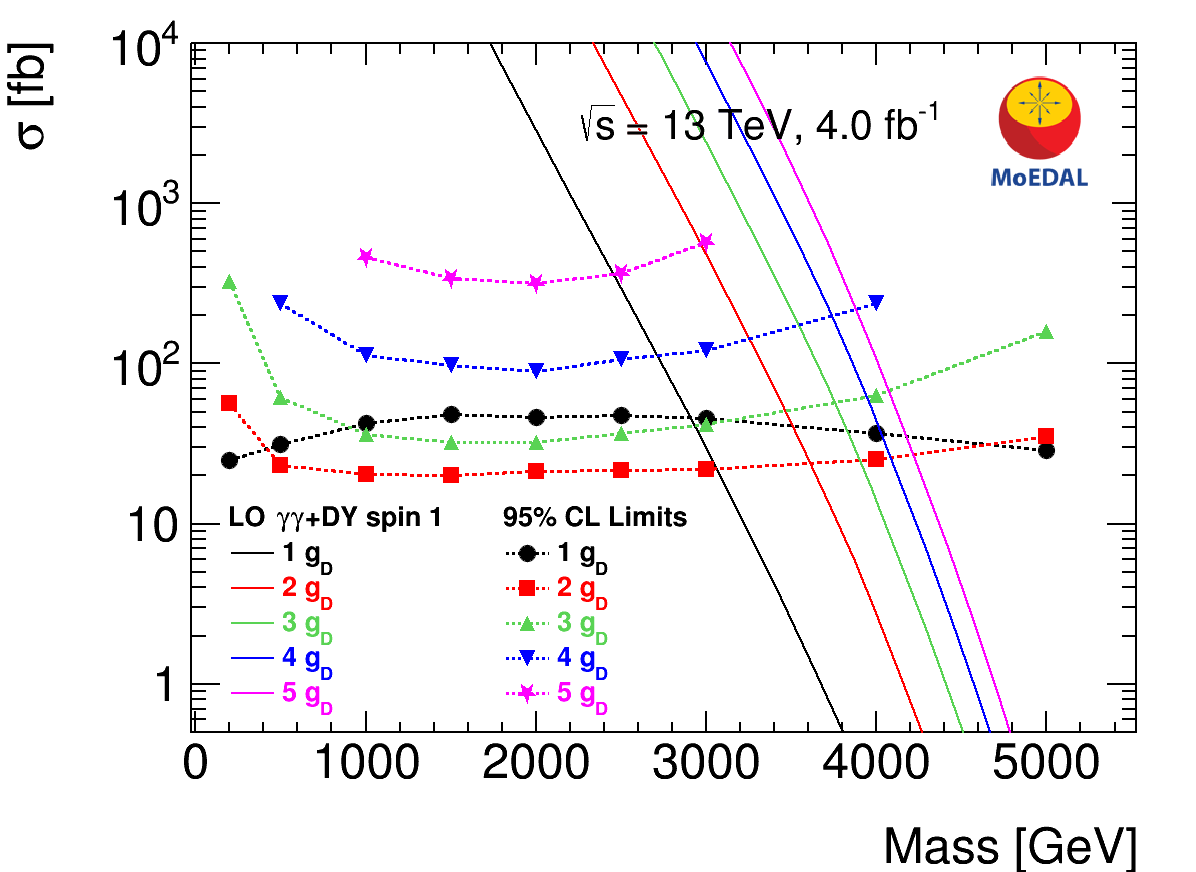}  
    \includegraphics[width=0.495\linewidth]{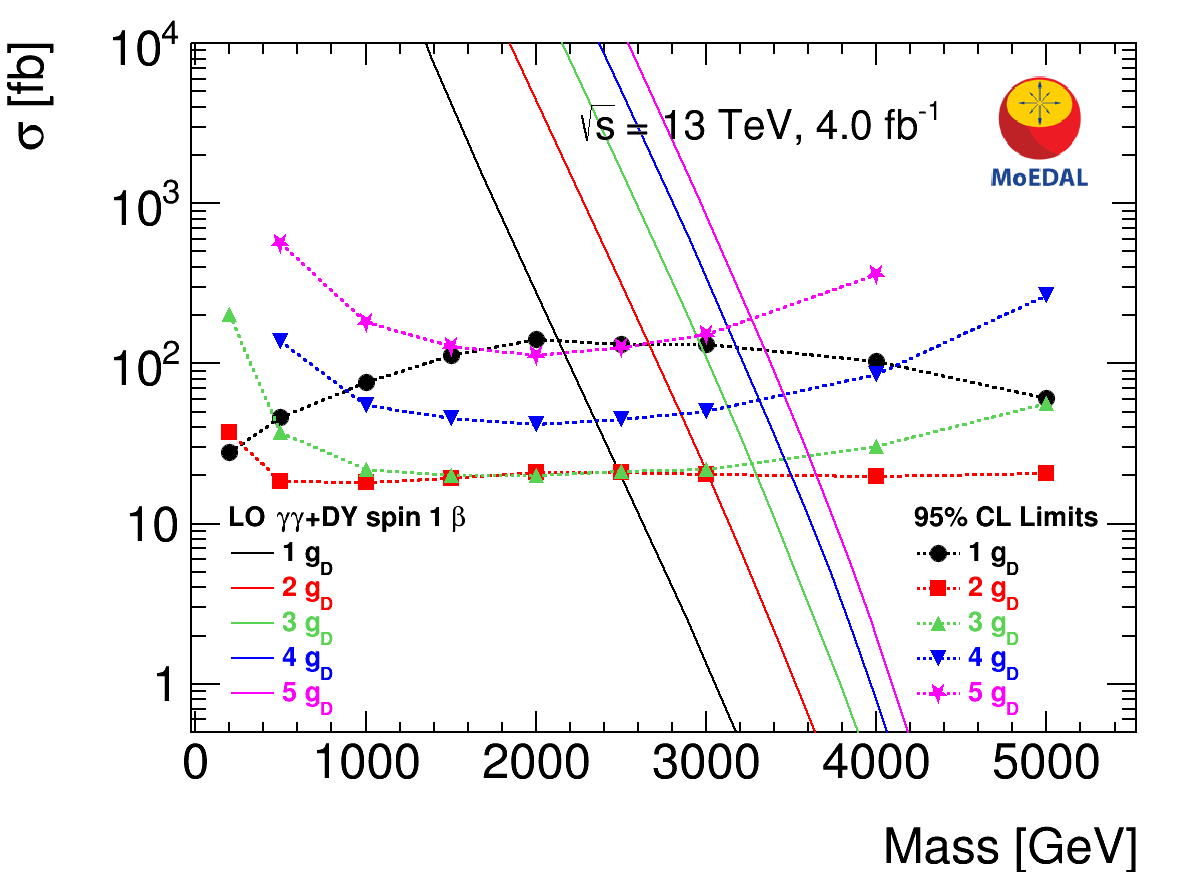}  
    \end{center}
  \caption{\label{fig:limits} Production cross-section upper limits at 95\% C.L.\ for the combined DY and \pf monopole pair production model with $\beta$-independent (left) and $\beta$-dependent (right) couplings in 13~\tev $pp$ collisions as a function of mass for spin-0 (top), spin-\half (middle), and spin-1 monopoles (bottom). The colors correspond to different monopole charges. The solid lines are cross-section calculations at leading order.}
\end{figure*}

The behavior of the acceptance as a function of mass has two contributions: the mass dependence of the kinematic distributions, and the energy loss, which decreases as the monopole slows down. For monopoles with $|g|=\gd$, acceptance losses predominantly come from punching through the trapping volume, resulting in the acceptance reaching a maximum of 3.8\% at low masses (high energy loss) and at high masses (low initial kinetic energy), having a minimum at around 3~\tev. The reverse is true for monopoles with $|g|>\gd$ that predominantly stop in the upstream material and for which the acceptance is highest (up to 4.5\% for $|g|=2\gd$, 4\% for $|g|=3\gd$, and 4\% for $|g|=4\gd$) for intermediate masses (around 2~\tev). The acceptance remains below 0.1\% over the whole mass range for monopoles carrying a charge of $6\gd$ or higher because they cannot be produced with sufficient energy to traverse the material upstream of the trapping volume. In this case, the systematic uncertainties become too large and the interpretation ceases to be meaningful. The dominant source of systematic uncertainties comes from the estimated amount of material in the \geant geometry description, yielding a relative uncertainty of $\sim10\%$ for 1\gd monopoles~\cite{MoEDAL:2016jlb}. This uncertainty increases with the magnetic charge reaching a point (at $6\gd$) where it is too large for the analysis to be meaningful. The spin dependence is solely due to the different event kinematics. The reader is referred to the Supplemental Material for more details on kinematic distributions and acceptances~\cite{supplemental}.

Production cross-section upper limits at 95\% confidence level (C.L.) for combined Drell-Yan and photon-fusion monopole production with the two coupling hypotheses ($\beta$-independent, $\beta$-dependent) and three spin hypotheses (0, \half, 1) are shown in Fig.~\ref{fig:limits}. They are extracted from the knowledge of the acceptance estimates and their uncertainties; the delivered integrated luminosity 4.0~\ifb, measured at a precision of 4\%~\cite{Aaij:2018okq}, corresponding to the 2015--2017 exposure to 13~\tev $pp$ collisions; the expectation of strong binding to aluminum nuclei~\cite{Milton:2006cp} of monopoles with velocity $\beta \le 10^{-3}$; and the nonobservation of magnetic charge $\ge\gd$ inside the trapping detector samples. Acceptance loss is dominated by monopoles punching through the trapping volume for $|g|=\gd$ while it is dominated by stopping in upstream material for higher charges, explaining the shape difference. Analogous limits considering DY production only are given in the Supplemental Material~\cite{supplemental} to facilitate comparison with previous MoEDAL~\cite{MoEDAL:2016jlb,Acharya:2016ukt,Acharya:2017cio} and ATLAS~\cite{Aad:2012qi,Aad:2015kta} results.

Production cross sections computed at leading order are shown as solid lines in Fig.~\ref{fig:limits}. Using these cross sections and the limits set by the search, indicative mass limits are extracted and reported in Table~\ref{tab:masslimits} for magnetic charges up to $5\gd$. No mass limit is given for the spin-0 and spin-\half $5\gd$ monopole with standard pointlike coupling, because in this case the low acceptance at small mass does not allow MoEDAL to exclude the full range down to the mass limit set at the Tevatron of around 400~\gev for DY models~\cite{Kalbfleisch:2003yt}. We note that these mass limits are only indicative, since they rely upon cross sections computed (at leading order) using perturbative field theory when the monopole-photon coupling is too large to justify such an approach.

\begin{table}[tb]
\caption{\label{tab:masslimits} 95\% C.L.\ mass limits in models of spin-0, \mbox{spin-\half} and spin-1 monopole pair direct production in LHC $pp$ collisions. The present results are interpreted for Drell-Yan and combined DY and photon-fusion production with both $\beta$-independent and $\beta$-dependent couplings.  }
\begin{ruledtabular}
\begin{tabular}{lcrrrrr}
Process / & \multirow{2}{*}{Spin}  & \multicolumn{5}{c}{Magnetic charge [$g_{\rm D}$] } \\
coupling &  & 1 & 2 & 3 & 4  & 5 \\
\colrule
& & \multicolumn{5}{c}{95\% C.L.\ mass limits [GeV]  } \\
\colrule
 DY  & 0                           & 790      & 1150    & 1210  & 1130    & --   \\
 DY  & \half                & 1320     & 1730    & 1770  & 1640   & --    \\
 DY  & 1                          & 1400     & 1840    & 1950  & 1910   & 1800  \\
 DY $\beta$-dep.   & 0          & 670      & 1010     & 1080   & 1040    & 900   \\
 DY $\beta$-dep.  & \half & 1050      & 1450    & 1530  & 1450   & --  \\
 DY $\beta$-dep.  & 1           & 1220      & 1680    & 1790  & 1780   & 1710  \\
\colrule
 DY+\pf & 0                           &  2190  &  2930 & 3120 &  3090  & --   \\
 DY+\pf & \half                & 2420  &  3180 & 3360 &  3340 & -- \\
 DY+\pf & 1                           & 2920 & 3620 & 3750 & 3740  & --  \\
 DY+\pf $\beta$-dep. & 0               &  1500  & 2300 & 2590 &  2640 & --   \\
 DY+\pf $\beta$-dep. & \half  &   1760  &  2610  & 2870  & 2940  & 2900  \\
 DY+\pf $\beta$-dep. & 1         &  2120 & 3010 & 3270 & 3300 & 3270  \\
\end{tabular}
\end{ruledtabular}
\end{table}


In summary, the aluminum elements of the MoEDAL trapping detector exposed to 13~\tev LHC collisions during the period 2015--2017 were scanned using a SQUID-based magnetometer to search for the presence of trapped magnetic charge. No candidates survived our scanning procedure and cross-section upper limits as low as 11~fb were set, improving previous limits of 40~fb also set by MoEDAL~\cite{Acharya:2017cio}. We considered the combined photon-fusion and Drell-Yan monopole-pair direct production mechanisms; the former process for the first time at the LHC. Consequently, mass limits in the range 1500--3750~\gev were set for magnetic charges up to $5\gd$ for monopoles of spins~0, \half and~1 ---the strongest to date at a collider experiment~\cite{Tanabashi:2018oca} for charges ranging from two to five times the Dirac charge. For a comparison, previous DY mass limits set by MoEDAL at 13~\tev ranged from 450 to 1790~\gev~\cite{Acharya:2017cio}.



We thank CERN for the very successful operation of the LHC, as well as the support staff from our institutions without whom MoEDAL could not be operated efficiently. We acknowledge the invaluable assistance of members of the LHCb Collaboration, in particular Guy Wilkinson, Rolf Lindner, Eric Thomas, and Gloria Corti. We thank Lucian Harland-Lang for discussions on the SuperChic event generator~\cite{Harland-Lang:2018iur} and on heavy-ion collisions. Computing support was provided by the GridPP Collaboration~\cite{Faulkner:2006px,Britton:2009ser}, in particular from the Queen Mary University of London and Liverpool grid sites. This work was supported by Grant No.\ PP00P2\_150583 of the Swiss National Science Foundation; by the UK Science and Technology Facilities Council (STFC), via the research Grants No.\ ST/L000326/1, No.\ ST/L00044X/1, No.\ ST/N00101X/1, and No.\ ST/P000258/1; by the Generalitat Valenciana via a special grant for MoEDAL and via the Project No.\ PROMETEO-II/2017/033; by the Spanish Ministry of Science, Innovation and Universities (MICIU), via the Grants No.\ FPA2015-65652-C4-1-R, No.\ FPA2016-77177-C2-1-P, No.\ FPA2017-85985-P, and No.\ FPA2017-84543-P; by the Severo Ochoa Excellence Centre Project No.\ SEV-2014-0398; by a 2017 Leonardo Grant for Researchers and Cultural Creators, BBVA Foundation; by the Physics Department of King's College London; by a Natural Science and Engineering Research Council of Canada via a project grant; by the V-P Research of the University of Alberta; by the Provost of the University of Alberta; by UEFISCDI (Romania); by the INFN (Italy); and by the Estonian Research Council via a Mobilitas Plus grant MOBTT5.

\bibliography{MMT2018.bib}

\end{document}
